\input{aipcheck}

\documentclass[
    ,final            % use final for the camera ready runs
  ]
  {aipproc}

\layoutstyle{8x11single}

\begin{document}

\title[Self/anti-self charge conjugate]{Self/anti-self charge conjugate states in the helicity basis}


\author{Valeriy V. Dvoeglazov}{
  address={UAF, Universidad de Zacatecas, M\'exico}, 
email={vdvoeglazov@yahoo.com.mx},
homepage={http://fisica.uaz.edu.mx/~valeri/}
}

\begin{abstract}
 We construct self/anti-self charge conjugate (Majorana-like) states for the $(1/2,0)\oplus (0,1/2)$
representation of the Lorentz group, and their analogs for higher spins within the quantum field theory. The problem of the basis rotations and that of the selection of phases in the Dirac-like and Majorana-like field operators are considered. The discrete symmetries properties (P, C, T) are studied. Particular attention has been paid to the question of (anti)commutation of the Charge conjugation operator and the Parity in the helicity basis. 
Dynamical equations have also been presented. 
In the $(1/2,0)\oplus (0,1/2)$ representation 
they obey the Dirac-like equation with eight components, which has been first introduced by Markov.
Thus, the Fock space for corresponding quantum fields is doubled (as shown by Ziino).
The chirality and the helicity 
(two concepts which are frequently confused in the literature) for 
Dirac and Majorana states have been discussed. 
\end{abstract}

%\date{\today}

\maketitle

%%%%%%%%%%%%%%%%%%%%%%%%%%%%%%%%%%%%%%%%%%%%
%% MAINMATTER
%%%%%%%%%%%%%%%%%%%%%%%%%%%%%%%%%%%%%%%%%%%%

The {\it self/anti-self} charge-conjugate 4-spinors
have been introduced 
in, e.~g., Refs.~\cite{Majorana,Bilenky,Ziino,Ahluwalia2}
in the coordinate representation and in the momentum representation. 
Later, these spinors have been studied in 
Refs.~\cite{Lounesto,Dvoeglazov1a,Dvoeglazov2,Kirchbach,Rocha1}.
The authors found corresponding dynamical equations, gauge transformations 
and other specific features of them. On using $C=  -e^{i\theta} \gamma^2 {\cal K}\,
\,$\footnote{ {\cal K}
is the complex conjugation operation. $\gamma^2$ is the matrix of the set of $\gamma$ matrices 
in the Weyl basis.} 
we  define the {\it self/anti-self} charge-conjugate 4-spinors 
in the momentum space
$C\lambda^{S,A} ({\bf p}) = \pm \lambda^{S,A} ({\bf p})\,,\,
C\rho^{S,A} ({\bf p}) = \pm \rho^{S,A} ({\bf p})$.
Such definitions of 4-spinors differ, of course, from the original Majorana definition in the x-representation
$\nu (x) = \frac{1}{\sqrt{2}} (\Psi_{Dirac} (x) + \Psi_{Dirac}^c (x))$,
$C \nu (x) = \nu (x)$ that represents the positive real $C-$ parity only. Nevertheless,
both definitions are connected each other, that permits to call them ``Majorana-like". 
In the spinorial basis with the appropriate normalization (``mass dimension")
the explicit forms of the 4-spinors of the second kind  $\lambda^{S,A}_{\uparrow\downarrow}
({\bf p})$ (and the corresponding expressions for $\rho^{S,A}_{\uparrow\downarrow} ({\bf p})$)
have been given, e.~g., in Ref.~\cite{Dvoeglazov1a}:
\begin{eqnarray}
\hspace{-10mm}\lambda^S_\uparrow ({\bf p}) &=& \frac{1}{2\sqrt{E+m}}
\pmatrix{ip_l\cr i (p^- +m)\cr p^- +m\cr -p_r},\,\,
\lambda^S_\downarrow ({\bf p})= \frac{1}{2\sqrt{E+m}}
\pmatrix{-i (p^+ +m)\cr -ip_r\cr -p_l\cr (p^+ +m)},\\
\hspace{-10mm}\lambda^A_\uparrow ({\bf p}) &=& \frac{1}{2\sqrt{E+m}}
\pmatrix{-ip_l\cr -i(p^- +m)\cr (p^- +m)\cr -p_r },\,\,
\lambda^A_\downarrow ({\bf p}) = \frac{1}{2\sqrt{E+m}}
\pmatrix{i(p^+ +m)\cr ip_r\cr -p_l\cr (p^+ +m)},
\end{eqnarray}
and
\begin{eqnarray}
\hspace{-10mm}\rho^S_\uparrow ({\bf p}) &=& \frac{1}{2\sqrt{E+m}}
\pmatrix{p^+ +m\cr p_r\cr ip_l\cr -i(p^+ +m)},\,\,
\rho^S_\downarrow ({\bf p}) = \frac{1}{2\sqrt{E+m}}
\pmatrix{p_l\cr (p^- +m)\cr i(p^- +m)\cr -ip_r},\\
\hspace{-10mm}\rho^A_\uparrow ({\bf p}) &=& \frac{1}{2\sqrt{E+m}}
\pmatrix{p^+ +m\cr p_r\cr -ip_l\cr i (p^+ +m)},\,\,
\rho^A_\downarrow ({\bf p}) = \frac{1}{2\sqrt{E+m}}
\pmatrix{p_l\cr (p^- +m)\cr -i(p^- +m)\cr ip_r},
\end{eqnarray}
where $p_{r,l}= p_x \pm ip_y$ and $p^\pm = E\pm p_z$.
As claimed in~\cite{Ahluwalia2}, $\lambda-$ and $\rho-$ 4-spinors are {\it not} the eigenspinors of 
the helicity ($h$). 
Moreover, 
$\lambda-$ and $\rho-$ are {\it not} the eigenspinors of the parity, as opposed to the Dirac case.
The authors  claimed in the cited works that indices $\uparrow\downarrow$ should be referred to the chiral helicity quantum number introduced 
in the 60s, $\eta=-\gamma^5 h$, Ref.~\cite{SenGupta}.
The normalizations of the spinors have also been given in the previous works.

We can introduce the quantum fields $\Psi (x)$, Ref.~\cite{BLP}, now composed of the $\lambda-$ and $\rho-$ spinors.
For instance, $\lambda^S ({\bf p}) \exp (-i p\cdot x)$ and $\rho^A ({\bf p}) \exp (-ip\cdot x)$ correspond to the 
``positive-energy solutions", and $\lambda^A ({\bf p}) \exp (+i p\cdot x)$ and 
$\rho^S ({\bf p})\exp (+i p\cdot x)$,
to ``the negative-energy solutions".
 In this case, the dynamical coordinate-space equations are:
\begin{eqnarray}
i \gamma^\mu \partial_\mu \lambda^S (x) - m \rho^A (x) &=& 0 \,,\,
i \gamma^\mu \partial_\mu \rho^A (x) - m \lambda^S (x) = 0 \,,
\label{12}\\
i \gamma^\mu \partial_\mu \lambda^A (x) + m \rho^S (x) &=& 0\,,\,
i \gamma^\mu \partial_\mu \rho^S (x) + m \lambda^A (x) = 0\,.
\label{14}
\end{eqnarray}
These are {\it not} the 4-component Dirac equations.
Similar formulations have been presented by M. Markov~\cite{Markov}, and by
A. Barut and G. Ziino~\cite{Ziino}. The group-theoretical basis for such doubling has been given
in the papers by Gelfand, Tsetlin and Sokolik~\cite{Gelfand}, who first presented 
the theory in the 2-dimensional representation of the inversion group (later called as the Bargmann-Wightman-Wigner-type quantum field theory).
The Lagrangian contains all 4-spinors ($\lambda^{S,A}$ and $\rho^{S,A}$).
The connection with the Dirac spinors has been found~\cite{Dvoeglazov1a,Kirchbach}. 

It was shown~\cite{Dvoeglazov1a} that the covariant derivative (and, hence, the
 interaction) can be introduced in this construct in the following way
$\partial_\mu \rightarrow \nabla_\mu = \partial_\mu - ig \L^5 A_\mu$,
where $\L^5 = \mbox{diag} (\gamma^5 \quad -\gamma^5)$.
Next, the Majorana-like field operator
($b^\dagger \equiv a^\dagger$)\footnote{The creation operators for a particle and an anti-particle are identical here.} admits additional phase (and, in general,
normalization) transformations $\nu^{ML\,\,\prime}
(x^\mu) = \left [ c_0 + i({\bf \tau}\cdot  {\bf c}) \right
]\nu^{ML\,\,\dagger} (x^\mu)$,
where $c_\alpha$ are
arbitrary parameters, $\tau-$ matrices are the analogs of the Pauli matrices, which are defined over the field of the $2 \times 2 $ matrices. 
Recently, the interest to these models raised again~\cite{Rocha1,Rocha2}.
We showed that the helicity, chiral helicity and chirality
operators are connected by unitary transformations, see below. The first one is 
\begin{eqnarray}
{\cal U}_1 =\pmatrix{1& p_l/(p+p_3)\cr
-p_r/(p+p_3)&1\cr},\,\,U_1 =\pmatrix{{\cal U}_1 &0\cr
0& {\cal U}_1\cr}\,,\,\,
U_1 \hat h  U_1^{-1} = 
\vert \frac{{\bf n}}{2} \vert \pmatrix{\sigma_3 &0\cr
0&\sigma_3 },\nonumber
\hspace*{-10mm}\\
\end{eqnarray}
with subsequent applications of the matrices
\begin{eqnarray}
U_2=\pmatrix{1&0&0&0\cr
0&0&0&1\cr
0&0&1&0\cr
0&1&0&0\cr },\,\,
U_3=\pmatrix{1&0&0&0\cr
0&0&1&0\cr
0&1&0&0\cr
0&0&0&1\cr }.
\end{eqnarray}
The concept of the doubling of the Fock space has been
developed in Ziino works (cf.~\cite{Gelfand,Dvoeglazov5}). 

Several formalisms have been used for higher spin fields, e.~g.,~\cite{BW, Weinberg}.
Ahluwalia {\it et al.} write~\cite{Ahluwalia2}:
``For spin-$1$ ... the requirement of self/anti-self charge
conjugacy {\it cannot} be satisfied. That is, there does not exist a
$\zeta$ [the phase factors between right- and left- 3-``spinors"] that can satisfy the spin-$1$ ... requirement
$S^c_{[1]}
\,\lambda(p^\mu)\,=\,\pm\,\lambda(p^\mu)\,\,,\quad S^c_{[1]}
\,\rho(p^\mu)\,=\,\pm\,\rho(p^\mu)$\,".
This is due to the fact that $C^2 = -1$ within this definition of the charge conjugation
operator. ``We find, however, that the requirement of self/anti-self
conjugacy under charge conjugation can be replaced by the requirement
of self/anti-self
conjugacy under the operation of $\Gamma^5\,S^c_{[1]}$" [precisely, which was used by Weinberg in Ref.~\cite{Weinberg}
due to the different choice of the equation for the negative-frequency 6-``bispinors"]. 
The covariant equations for the $\lambda-$ and $\rho-$ objects in the $(1,0)\oplus (0,1)$
representation have been obtained~in~Ref.~\cite{Dvoeglazov1a}. For instance,\footnote{$\gamma_{\mu\nu}$ are the covariantly-defined matrices of this representation space.}
\begin{eqnarray}
&&\hspace*{-5mm}\gamma_{\mu\nu} p^\mu p^\nu \lambda^S_{\uparrow\downarrow\rightarrow}
(p^\mu) - m^2 \rho^S_{\uparrow\downarrow\rightarrow} (p^\mu)=0,
\gamma_{\mu\nu} p^\mu p^\nu \lambda^A_{\uparrow\downarrow\rightarrow}
(p^\mu) - m^2 \rho^A_{\uparrow\downarrow\rightarrow} (p^\mu)=0,\\
&&\hspace*{-5mm}\gamma_{\mu\nu} p^\mu p^\nu \rho^S_{\uparrow\downarrow\rightarrow} (p^\mu)
- m^2 \lambda^S_{\uparrow\downarrow\rightarrow} (p^\mu)=0,
\gamma_{\mu\nu} p^\mu p^\nu \rho^A_{\uparrow\downarrow\rightarrow} (p^\mu)
- m^2 \lambda^A_{\uparrow\downarrow\rightarrow} (p^\mu)=0\,.
\end{eqnarray}

Next, R. da Rocha {\it et al.} write~\cite{Rocha2} that in the Dirac case
$\{C,P\}_+=0$. On  the other hand,
it follows that $[C,P]_-=0$ when acting on the Majorana-like states. In the previous works of 
the 50s-60s, see, e.~g., Ref.~\cite{NigamFoldy},
it is the latter case which has been attributed to the $Q=0$ eigenvalues (the truly neutral particles).
You may compare these results with those of~\cite{Ahluwalia2,Dvoeglazov2,Dvoeglazov4},
where the same statements have been done on the quantum-field level even 
at the earlier time comparing with Ref.~\cite{Rocha2}. 
The acronym "ELKO" is the synonym for the self/anti-self charge conjugated states (the Majorana-like spinors).
It is easy to find the correspondence between ``the new notation", Refs.~\cite{Ahluwalia3,Rocha2}
and the previous one. Namely, $\lambda^{S,A}_{\uparrow} \rightarrow \lambda^{S,A}_{-,+}$,
$\lambda^{S,A}_{\downarrow} \rightarrow \lambda^{S,A}_{+,-}$. However, the difference
is also in the choice of the basis for the 2-spinors. As in Ref.~\cite{Dvoeglazov3},
Ahluwalia, Grumiller and da Rocha have chosen the well-known helicity basis (cf.~\cite{Varshalovich,Dv-ff}).
I have shown that the helicity-basis 4-spinors
satisfy the same Dirac equation, the parity matrix can be defined in the similar fashion 
as in the spinorial basis, but 
the helicity-basis 4-spinors are {\it not} the eigenspinors of the parity,
Ref.~\cite{BLP}. 

Usually, the 2-spinors  are parametrized
in the following way, cf.~\cite{Varshalovich,Dv-ff}, in the helicity 
basis:\footnote{Nevertheless, this parametrization
immediately implies some controversies in the properties of the overall and relative phases
with respect to the space inversion.}
\begin{eqnarray}
\phi_L^+ =Ne^{i\theta_1} \pmatrix{\cos (\vartheta/2) e^{-i\varphi/2}\cr
\sin (\vartheta/2) e^{+i\varphi/2}},\,\,
\phi_L^+ =Ne^{i\theta_2} \pmatrix{\sin (\vartheta/2) e^{-i\varphi/2}\cr
-\cos (\vartheta/2) e^{+i\varphi/2}}.\,
\end{eqnarray}
Next, their analogs for spin 1 are:
\begin{eqnarray}
\phi_L^{+} =
N\,e^{i\theta_1}\,\pmatrix{{1\over 2} (1+\cos\vartheta) e^{-i\varphi}\cr
\sqrt{{1\over 2}} \sin\vartheta\cr
{1\over 2} (1-\cos\vartheta) e^{+i\varphi}\cr}&&, \,
\phi_L^{-} = N\,e^{i\theta_2}\,\pmatrix{-{1\over 2} (1-\cos\vartheta)
e^{-i\varphi}\cr
\sqrt{{1\over 2}} \sin\vartheta\cr
-{1\over 2} (1+\cos\vartheta) e^{+i\varphi}\cr},\\
&&\hspace*{-25mm}\phi_L^{0} =  N\,e^{i\theta_0}\,
\pmatrix{-\sqrt{{1\over 2}}\sin \vartheta \,e^{-i\varphi}\cr
\cos\vartheta\cr
\sqrt{{1\over 2}}\sin \vartheta \,e^{+i\varphi}\cr}\,,
\end{eqnarray}
In this basis, the parity transformation  for 2-spinors
leads to the properties (provided that the overall phases do {\it not} transform):
\begin{eqnarray}
R \phi_L^- ({\bf 0}) &=& -i e^{i(\theta_2 -\theta_1)}\phi_L^+ ({\bf 0})\,,\,
R \phi_L^+ ({\bf 0}) = -i e^{i(\theta_1 -\theta_2)}\phi_L^- ({\bf 0}),\label{helic1}\\
R \Theta (\phi_L^- ({\bf 0}))^\ast &=& -i e^{-2i\theta_2}\phi_L^- ({\bf 0})\,,\,
R \Theta (\phi_L^+ ({\bf 0}))^\ast =  +i e^{-2i\theta_1}\phi_L^+ ({\bf 0})\label{helic2},
\end{eqnarray}
$\Theta=-i\sigma_y$ is the Wigner matrix.
In the $(1,0)\oplus (0,1)$ representation the situation is similar (see the formulas (31) in Ref.~\cite{Dv-ff}). The analogs of (\ref{helic1},\ref{helic2}) are:
\begin{eqnarray}
R \phi_L^- ({\bf 0}) &=& + e^{i(\theta_2 -\theta_1)}\phi_L^+ ({\bf 0})\,,\,
R \phi_L^+ ({\bf 0}) = + e^{i(\theta_1 -\theta_2)}\phi_L^- ({\bf 0}),\label{helic11}\\
R \Theta (\phi_L^- ({\bf 0}))^\ast &=& - e^{-2i\theta_2}\phi_L^- ({\bf 0})\,,\,
R \Theta (\phi_L^+ ({\bf 0}))^\ast =  - e^{-2i\theta_1}\phi_L^+ ({\bf 0}),\label{helic21}\\
R \phi_L^0 ({\bf 0}) &=& - \phi_L^0 ({\bf 0})\,,\,
R \Theta (\phi_L^0 ({\bf 0}))^\ast = + e^{-2i\theta_0}\phi_L^0 ({\bf 0}).\label{helic31}
\end{eqnarray}
This  opposes  to the spinorial basis, where, for instance:
$R \phi_L^{\pm} ({\bf 0}) =\phi_L^{\pm} ({\bf 0})$. 

Correspondingly, if we choose the Parity operator as $P=i\gamma^0 R$, cf.~\cite{Majorana,Ahluwalia3},
we have in the $(1/2,0)\oplus (0,1/2)$ representation:
\begin{eqnarray}
P\lambda^S_\uparrow ({\bf 0}) &=& -i \cos (2\theta_1) \lambda^A_\uparrow ({\bf 0})- \sin(2\theta_1) \lambda^S_\uparrow ({\bf 0})\,,\\
P\lambda^S_\downarrow ({\bf 0}) &=& i \cos (2\theta_2) \lambda^A_\downarrow ({\bf 0})+ \sin(2\theta_2) \lambda^S_\downarrow ({\bf 0})\,,
\end{eqnarray}
and, in the $(1,0)\oplus (0,1)$ representation:
\begin{eqnarray}
P\lambda^S_\uparrow ({\bf 0}) &=& -i \cos (2\theta_1) \lambda^S_\uparrow ({\bf 0})- \sin(2\theta_1) \lambda^A_\uparrow ({\bf 0})\,,\\
P\lambda^S_\rightarrow ({\bf 0}) &=& i \cos (2\theta_0) \lambda^A_\rightarrow ({\bf 0})+ \sin(2\theta_0) \lambda^S_\rightarrow ({\bf 0})\,\\
P\lambda^S_\downarrow ({\bf 0}) &=& -i \cos (2\theta_2) \lambda^S_\downarrow ({\bf 0})- \sin(2\theta_2) \lambda^A_\downarrow ({\bf 0})\,.
\end{eqnarray}
Analogous equations can be derived for $\lambda^{A}_{\uparrow\downarrow}$.

Further calculations are straightforward in both  $(1/2,0)\oplus (0,1/2)$ and $(1,0)\oplus (0,1)$ representations. And, indeed,  they can lead to $[C,P]_- =0$ when acting on the "ELKO" states,
provided that we take into account the phase factor in the definition of the $P-$ operator. 

We presented a review of the formalism in the momentum-space Majorana-like
particles in the $(S,0)\oplus (0,S)$ representation of the Lorentz Group. The $\lambda-$ and $\rho-$
4-spinors  satisfy the 8-
component analogue of the Dirac equation. Apart, they have different gauge transformations
comparing with the usual Dirac 4-spinors. Their helicity, chirality and chiral helicity properties
have been investigated in detail. 
At the same time, we showed that the Majorana-like 4-spinors can be obtained
by the rotation of the spin-parity basis. Meanwhile, several authors have claimed that the
physical results would be different on using calculations with these Majorana-like spinors.
However, several statements made by other researchers concerning with chirality, helicity, chiral
helicity, and $C$, $P$ (anti)commutation should not be considered to be true until the time when experiments confirm them.
Usually, it is considered that the rotations (unitary transformations) have {\it no} any physical
consequences on the level of the Lorentz-covariant theories.
Next, we discussed the $[C,P]_{\pm}=0$ dilemma for neutral and charged particles
on using the analysis of the basis rotations and phases.

%%%%%%%%%%%%%%%%%%%%%%%%%%%%%%%%%%%%%%%%%%%%%%%%
%% The bibliography can be prepared using the BibTeX program or
%% manually.
%%
%% The code below assumes that BibTeX is used.  If the bibliography is
%% produced without BibTeX comment out the following lines and see the
%% aipguide.pdf for further information.
%%
%% For your convenience a manually coded example is appended
%% after the \end{document}
%%%%%%%%%%%%%%%%%%%%%%%%%%%%%%%%%%%%%%%%%%%%%%%%

%%%%%%%%%%%%%%%%%%%%%%%%%%%%%%%%%%%%%%%%%%%%%%%%
%% You may have to change the BibTeX style below, depending on your
%% setup or preferences.
%%
%%
%% For The AIP proceedings layouts use either
%%%%%%%%%%%%%%%%%%%%%%%%%%%%%%%%%%%%%%%%%%%%

\bibliographystyle{aipproc}   % if natbib is available
%\bibliographystyle{aipprocl} % if natbib is missing

%%%%%%%%%%%%%%%%%%%%%%%%%%%%%%%%%%%%%%%%%%%
%% You probably want to use your own bibtex database here
%%%%%%%%%%%%%%%%%%%%%%%%%%%%%%%%%%%%%%%%%%%
%\bibliography{sample}

%%%%%%%%%%%%%%%%%%%%%%%%%%%%%%%%%%%%%%%%%%%
%% Just a reminder that you may have to run bibtex
%% All of it up to \end{document} can be removed
%% if you don't like the warning.
%%%%%%%%%%%%%%%%%%%%%%%%%%%%%%%%%%%%%%%%%%%
%\IfFileExists{\jobname.bbl}{}
% {\typeout{}
%  \typeout{******************************************}
%  \typeout{** Please run "bibtex \jobname" to optain}
%  \typeout{** the bibliography and then re-run LaTeX}
%  \typeout{** twice to fix the references!}
%  \typeout{******************************************}
%  \typeout{}
% }

%%%%%%%%%%%%%%%%%%%%%%%%%%%%%%%%%%%%%%%%%%%
%% The following lines show an example how to produce a bibliography
%% without the help of the BibTeX program. This could be used instead
%% of the above.
%%%%%%%%%%%%%%%%%%%%%%%%%%%%%%%%%%%%%%%%%%%

\end{document}